\documentstyle[12pt]{article}

\begin{document}
\renewcommand{\arraystretch}{1.5}
\begin{center}
\begin{Large}
{\bf Transport and Boundary Scattering in Confined\\
 Geometries: Analytical Results}\\
\end{Large}
\bigskip
R. A. Richardson and Franco Nori\footnote{to whom correspondence should be
addressed.}\\
{\em Department of Physics, University of Michigan, Ann Arbor MI 48109-1120}
\bigskip
\bigskip
\begin{large}
{\bf Abstract}
\end{large}
\end{center}

\noindent
\baselineskip 4.9ex
We utilize a geometric argument to determine the effects of boundary
scattering on the carrier mean-free path in samples of various cross
sections.  Analytic expressions for samples with rectangular and
circular cross sections are obtained.  We also outline a method for
incorporating these results into calculations of the thermal
conductivity.

\newpage

\noindent
\begin{center}
{\bf I. INTRODUCTION}
\end{center}

There has been considerable interest, in recent years, in the transport
properties of various materials in confined geometries (see, {\em
e.g.,} Refs.~[1]--[5]).  In many such applications, boundary
scattering  of carriers can be expected to play a significant role.
However, when efforts are made to calculate  transport quantities such
as the thermal conductivity for small samples, the effect of boundaries
is often handled in a fairly simplified fashion.  Boundary scattering
of phonons, for instance, is typically incorporated by the addition of
a constant term to the inverse scattering time for the phonon. The
purpose of this paper is to provide a concise expression for the effect
of boundaries on transport in small samples which improves upon this
basic approach, without invoking the full machinery of the Boltzmann
equation.
Expressions are obtained for axial transport in samples of circular and
rectangular cross section.  These results will be applicable to samples
which are small enough that the carrier mean-free path is on the order
of the sample dimensions but not so small that the carrier spectrum is
substantially modified from the bulk.  In other words, the sample
dimensions will be assumed to be much greater than the carrier
wavelength.

The method employed is based upon one proposed by Flik and Tien
\cite{flik} for the calculation of the size effect in thin films.
\ It is also related to a work put forward by Herring\cite{herring}
in which approximate expressions of the thermoelectric power for
some geometries are derived.
\ The Flik and Tien
method assumes that, for a carrier of a given frequency, a
characteristic bulk mean-free path, $l$, can be defined which is
determined by the other scatterers present in the sample.  The goal,
then, is to examine how this bulk value of $l$ is modified by the
presence of boundaries in the sample. The calculation utilizes the
concept of the exchange length $l_{ex}$ \cite{flik,tien}, which is
defined as the average distance normal to a plane that a carrier
travels after having been scattered within that plane. To be more
explicit, we consider a carrier that has undergone a scattering event
within a plane that is perpendicular to the direction of net transport
which will be referred to from here on as the positive $z$ direction.
We now allow the carrier to propagate to the point of its next
scattering event, which, in the bulk, is a distance $l$ away.  This
propagation is assumed to proceed with equal likelihood in all
directions. $l_{ex}$  is then defined as the average z-component of all
possible such propagation vectors, where the average is performed over
the hemisphere in the positive $z$ direction.  The bulk value of this
quantity, $l_{\infty}$ is determined by the expression \cite{flik},
 \begin{equation}
l_{\infty} =  \frac{1}{2 \pi} \int_{0}^{2\pi}
 d\phi \int_{0}^{\pi /2} d \theta \,l \sin\theta \, \cos\theta  = \frac{l}{2},
\end{equation}
where $\theta$ is the angle between the vector {\large \boldmath $l$}
and the $z$ axis and $\phi$ is the angle between the projection of
{\boldmath $l$} into the $xy$ plane and the positive $x$ axis.

In the following, we consider the scenario in which the mean-free path
is on the order of the sample dimensions.  For this case, some carriers
will strike the boundaries before traveling a full distance $l$ and the
exchange length will be correspondingly shorter. We will assume that
scattering at the boundaries is {\em diffuse}, which will be valid when
the carrier wavelength is smaller than the characteristic roughness
features of the sample surface.  We will also consider the sample to be
free of grain boundaries, though this calculation could perhaps also be
applied to samples whose grains have characteristic geometries which
match those investigated here. A geometric analysis will be used to
calculate the average exchange length for axial transport in narrow
samples (length is assumed infinite) of both circular and rectangular
cross section.
This quantity will then be related to an effective mean-free path,
$l_{\em eff}$, which can be used to calculate axial transport
coefficients. The calculations presented in the following sections are
somewhat lengthy, but we assure the reader that the method pursued,
particularly for the rectangular case, is considerably shorter than an
approach one might n\"{a}ively take.

\medskip
\noindent
\begin{center}
{\bf II. CALCULATION OF THE EXCHANGE LENGTH}\\
{\bf  A. Circular cross section}
\end{center}

In order to calculate the exchange length for a cylindrical sample, we
will initially assume that the excitation can originate with equal
likelihood anywhere within a given circular cross section of the
sample.  The average value of the exchange length in the sample, $
\tilde{l}_{ex} $, will then be obtained by averaging $l_{ex}$ (which is
itself an average over a hemisphere of solid angle) over the entire
cross section.
The geometry to be considered is shown in Fig.~1.  We consider an
excitation originating at some point a distance $\rho$ from the center
of the cross section of radius $R$ and propagating in a random
direction within the hemisphere of solid angle whose base is normal to
the positive $z$ direction. The quantity $\theta$ is defined as the
angle between the propagation vector, {\boldmath $l$}  and the z-axis,
and $\phi$ is the angle between the radius along which the origination
point  is located and the projection of  {\boldmath $l$}  into the
plane of the cross section. Note that  {\boldmath $l$}  may or may not
have length $l$, depending upon whether or not it is truncated by a
boundary.
We will also find it useful to define the quantity $s(\rho ,\phi)$ as
the length of the segment within the cross section plane which begins
at the origination point and extends to the sample boundary in the
direction of $\phi$ (see Fig.~1).  This will be the length of the
projection of  {\boldmath $l$}  into the plane of the cross section for
the case in which the carrier hits the wall before travelling a full
mean-free path, $l$.  The average exchange length will then be given by
the expression,
\begin{equation}
 \tilde{l}_{ex}  =\frac{2}{\pi R^2} \int_{0}^{R}
 d \rho   \rho  \int_{0}^{\pi}d\phi
 \int_{0}^{\pi/2}d\theta \sin{\theta} \,l_z( \rho ,\phi,\theta),
\end{equation}
where $l_z$ is the $z$ component of the propagation vector heading in
the $(\theta,\phi)$ direction and $\phi$ has only been integrated over
half its range for symmetry reasons.  In the above, an integration over
the polar angle $\xi$ within the cross section plane has already been
performed since $l_z$ does not depend on it.

We now need to consider the separate cases in which  {\boldmath $l$}
does and does not hit the wall.  For a given $ \rho $ and $\phi$,
{\boldmath $l$}  will hit within the range  $\theta^* < \theta < \pi/2$
where $\theta^*$, is given by $$ \theta^* = \sin^{-1}\left(\frac{s(
\rho ,\phi)}{l}\right).$$ The corresponding expressions for $l_z$ for
these two cases are given by
\begin{equation}
l_z( \rho ,\phi,\theta) = \left\{ \begin{array}{ll}
l \cos{\theta}  &  \mbox{ if $\; 0 < \theta < \theta^*$}\\
s( \rho ,\phi) \cot{\theta}  &  \mbox{ if $\; \theta^*< \theta < \pi/2$}.
\end{array} \right.
\end{equation}
We will also encounter situations, when $l < 2R$, where, for certain
ranges of $ \rho $ and $\phi$, $s( \rho ,\phi) > l$.  In these cases,
{\boldmath $l$}  {\em never} hits the wall for {\em any} $\theta$.  How
these ranges are determined will be examined more fully below.

We are now in a position to evaluate the $\theta$ integral.  We will
distinguish between the two cases described above, {\em i.e.}
{\boldmath $l$} {\em sometimes} hits the wall ($l_z$ is described by
Eq.~(3)) and,  {\boldmath $l$} {\em never} hits the wall ($l_z =
l\cos{\theta}$ for all $\theta$). These will be referred to hereafter
as the SH and NH regimes. The result of the $\theta$ integration is
\begin{equation}
\int_{0}^{\pi/2} d\theta \sin \theta \,l_z( \rho ,\phi,\theta)
 = \left\{ \begin{array}{ll}
l/2     &  \mbox{never hits (NH)}, \\
s( \rho ,\phi) -s( \rho ,\phi)^2/2l  &  \mbox{sometimes hits (SH).}
\end{array} \right.
\end{equation}

To proceed further, we must consider the relative magnitudes of $l$ and
$R$.  There are two relevant possibilities, $l \geq 2R$ and $l < 2R$.
We will consider the simpler of these two cases first,  $l \geq 2R$.
In this regime, one can see that, with regards to the $\theta$
integration, we will always ``sometimes hit" the wall, for all possible
values of $ \rho $ and $\phi$.  After performing the $\theta$
integration, the expression for  $ \tilde{l}_{ex} $ becomes
\begin{equation}
 \tilde{l}_{ex}  =\frac{2}{\pi R^2} \int_{0}^{R} d \rho
 \, \rho  \int_{0}^{\pi}d\phi \,
\left(s( \rho ,\phi) -\frac{s( \rho ,\phi)^2}{2l}\right)
\end{equation}
where, as can be determined from the geometry of the problem,
\begin{equation}
 s( \rho ,\phi) = (R^2- \rho ^2\sin^2\phi)^{1/2} -  \rho \cos{\phi}.
\end{equation}

It is possible to integrate the $-s^2/2l$ term in the order written
above ($\phi$ first), but it can be shown that the $s$ term must be
evaluated by performing the $ \rho $ integration first to achieve an
analytical solution. These operations yield the following result:
\begin{equation}
 \tilde{l}_{ex}  = \frac{8R}{3\pi} -\frac{R^2}{2l}
\mbox{\hspace*{1in} $l \geq 2R.$}
\end{equation}

The case where $l < 2R$ is somewhat more involved.  Here we must
distinguish between those regions of $( \rho ,\phi)$ space where we are
in the SH regime with regards to the $\theta$ integration and those
where we are in the NH regime.  The following results will be derived
upon the assumption that $l<R$.  The remaining case, $ R < l < 2R$, can
be shown, though it is not readily apparent, to yield results identical
to the $l<R$ case and we will therefore spare the reader their
derivation.
Assuming then, that  $l<R$, one can see that in the region where $ \rho
< R-l$,              {\boldmath $l$} never hits the wall, regardless of
$\phi$ and $\theta$.  For $ \rho  \geq R-l$, on the other hand, we can
be either in the SH or NH regime, depending upon the value of $ \rho $
and $\phi$.  It can be seen that, given a $ \rho $ greater than $R-l$,
the angle $\phi^*$ for which we cross over from the SH to the NH regime
is determined by the condition that $s(\rho,\phi^*) = l$. The geometry
for this situation is illustrated in Fig.~2.  Using this condition, it
can be shown that $\phi^* = \pi - \cos^{-1}(\frac{ \rho ^2 +l^2-R^2}{2l
\rho })$.  Summarizing these considerations for the $l< R$ case, we
arrive at the expression,
\begin{eqnarray}
\tilde{l}_{ex} & = & \frac{2}{\pi R^2}\left\{ \int_{R-l}^{R}
 d \rho   \rho  \left[\int_{0}^{\phi^*}d\phi \,
\left( s( \rho ,\phi) -\frac{s( \rho ,\phi)^2}{2l}\right)
 + \int_{\phi^*}^{\pi}
d\phi \,\frac{l}{2} \right] \right. \nonumber \\
  &   &   \left. + \int_{0}^{R-l} d \rho \rho \int_{0}^{\pi}d\phi \,
\frac{l}{2} \right\}.
\end{eqnarray}

The NH terms, where the $\phi$ integrand is $l/2$, as well as the term
involving $-s^2/2l$, can be evaluated as presented above.  However, the
``$s$ term", as given by

\begin{equation}
\mbox{``$s$ term"} = \frac{2}{\pi R^2}\int_{R-l}^{R}
d \rho   \rho  \int_{0}^{\phi^*}d\phi \,
 s( \rho ,\phi),
\end{equation}
can not be evaluated in the integration order shown, since the $\phi$
integral is elliptic.  So, as in the $l>2R$ case, we must reverse the
order of integration and approach the $\rho$ integral first.  This
procedure is less straightforward than the previous case, however, due
to the fact that the upper limit of the $\phi$ integral in Eq.~(9) is
$\rho$ dependent.

We must reconsider the problem from the perspective of integrating over
$\rho$ for constant $\phi$.  The relevant geometry for this case is
presented in Figs. 3a and b.  The goal is to determine the regions of
phase space where we are in the SH regime, as this is the only region
which involves the $s$ term that we seek to evaluate.  If we first
examine the different ranges of $\phi$, we can see that for large
values of this angle, we will never hit the boundary for any $\rho$ or
$\theta$.  The critical angle $\phi_2^*$ above which this will be the
case is determined by the condition that $s(R,\phi_2^*) =l$, (see
Fig.~3a) and can be shown to be given by $\phi_2^* = \pi -
\cos^{-1}(\frac{l}{2R})$. Now, directing our attention to the different
possible ranges of $\rho$,  one can see that for $\phi < \phi_2^*$, we
will be in the NH regime for small $\rho$ and will cross over to the SH
regime at a critical value $\rho^*$.  The determining condition for
this crossover will again be that
$s(\rho^*,\phi) =l$, (see Fig.~3b), which leads to the result $$\rho^*
= (R^2 - l^2 \sin^2 \phi)^{1/2} - l\cos \phi.$$ These various
considerations yield the proper expression for the $s$ term with the
integration order reversed,
\begin{equation}
\mbox{``s term"} = \frac{2}{\pi R^2} \int_{0}^{\phi_2^*}
d\phi \int_{\rho^*}^{R} d \rho \rho \:  s( \rho ,\phi),
\end{equation}
where $\phi_2^*$ and $\rho^*$ are given above.  This expression can now
be evaluated
(see Appendix A)
and, when combined with the results for
the other terms obtained from Eq.~(8), gives the total result for the
$l < 2R$ case:
\begin{eqnarray}
\tilde{l}_{ex}  & = & \frac{8R}{3\pi} -\frac{R^2}{2l} +
 \frac{l}{\pi}\cos^{-1}\left(\frac{l}{2R}\right) +
\frac{R^2}{2\pi l}\cos^{-1}\left(\frac{l^2}{2R^2} - 1 \right)
\mbox{\hspace{.5in}$l < 2R$}  \nonumber \\
   &  &  - \frac{(4R^2-l^2)^{1/2}}{\pi}\left(\frac{13}{12} +
 \frac{l^2}{24R^2} \right).
\end{eqnarray}
It is interesting to note that the $-s^2/2l$ term can {\em not} be
integrated in a straightforward fashion with the $\rho$ and $\phi$
integrations reversed, so both perspectives are necessary for a
complete solution to the problem.

\medskip
\noindent
\begin{center}
{\bf B. Rectangular Cross Section}
\end{center}

We now turn our attention to samples of rectangular cross section.  The
general expression for $\tilde{l}_{ex}$ is quite analogous to the
circular case and is given by
 \begin{equation}
 \tilde{l}_{ex}  =\frac{2}{\pi ab} \int_{0}^{a/2}
 dx \int_{0}^{b/2} dy  \int_{0}^{2\pi}d\phi
 \int_{0}^{\pi/2}d\theta \sin{\theta} \,l_z(x,y ,\phi,\theta),
\end{equation}
where $a$ and $b$ are the lengths of the two sides and the integration
is performed over only the bottom-left quarter of the cross section for
symmetry reasons.  The  bottom-left corner is chosen as the origin of
$x$ and $y$ and the $a$ side lies along the $x$ direction.
We will define $s(x,y,\phi)$ in a manner analogous to the previous case
{\em i.e.,} the distance from the origination point to the wall in the
$\phi$ direction, and  point out that the results of the $\theta$
integration can again be divided into  SH and NH regimes with the
results shown in Eq.~(4)  applicable here as well.  The expression for
$s$ will depend upon which boundary we are relating it to and, for the
bottom wall for instance, is given by
\begin{equation}
s(x,y,\phi) = y\sec\phi.
\end{equation}

The integrations over $\phi$, $x$, and $y$ require careful
consideration  to determine the regions of $(x,y,\phi)$ space where we
are in the SH or NH regimes.  We will find that, as before, the makeup
of these regions depends on the relative magnitudes of $l$, $a$, and
$b$.  We first examine the case when $l >\sqrt{a^2 +b^2}$. This is the
simplest scenario because we are in the SH regime for all $x$, $y$ and
$\phi$.  The geometry for this case is illustrated in Figs. 4 and 5.
For a given origination point $(x,y)$,  we  have to consider four
ranges of $\phi$ integration: $\phi_1 \rightarrow \phi_2$, $\phi_3
\rightarrow \phi_4$, $\phi_5 \rightarrow \phi_6$, and  $\phi_7
\rightarrow \phi_8$, corresponding to hitting the bottom, left, top and
right walls, respectively, (see Fig.~4).  For the bottom wall, $\phi$
is measured with respect to the vertical through the origination point,
with $\phi_1$ in the negative direction and $\phi_2$ in the positive
direction.  The counter-clockwise directional convention is chosen
because {\boldmath $l$} is directed into the plane of the paper.  To
avoid ambiguity in the evaluation of inverse tangents, which appear
frequently in the calculation, a new zero of $\phi$ is defined for each
side, with, for example, the zero with respect to the left side taken
as the horizontal to the left of the origination point.

We are now prepared to undertake the integration over the full range of
$\phi$.  This task can be considerably simplified, however, by taking
into account the symmetry of the problem.  This process is illustrated
schematically in Fig.~5. We begin by treating the contributions from
the top and bottom, hereafter referred to as the ``top/bottom" terms,
separately from the sides. We will find that the latter can be related
to the former by symmetry.  Next, we note that the integration over the
{\em bottom-left} quarter of our cross section of those $\phi$'s which
will place us in the region where scattering will occur from the {\em
top} boundary,
{\em i.e.,}  $\phi_5 \rightarrow \phi_6$, is equivalent, by a 180
degree rotation, to an integration over the {\em top-right} quarter of
the cross section of those $\phi$'s where scattering will occur from
the {\em bottom} of the sample, {\em i.e.,}  $\phi_1 \rightarrow
\phi_2$.  The part of our integral which accounts for scattering from
the top and bottom boundaries then becomes \begin{equation} \left[
\int_{0}^{a/2}dx\int_{0}^{b/2}dy + \int_{a/2}^{a}dx\int_{b/2}^{b}dy
\right] \int_{\phi_1(x,y)}^{\phi_2(x,y)}d\phi \, \left(s(x,y,\phi) -
\frac{s(x,y,\phi)^2}{2l}\right).
\end{equation}

The next simplification occurs as a result of an additional symmetry.
It can be shown that when the $\phi$ integral is evaluated at its lower
limit $\phi_1$, one obtains the same expression as when it is
evaluated at the upper limit, $\phi_2$, provided that the $\phi_1$
expression is subjected to a change of variables which changes the
region of integration in $(x,y)$ space.  Specifically, the appropriate
region is obtained by taking the original region and reflecting it
about the line $x = a/2$.  The benefit of this operation is that we
reduce an expression which involves two terms, the $\phi$ integral
evaluated at
 its upper and lower limits, to one involving only one term, simply by
changing the region of $(x,y)$ integration. In particular, we arrive at
an expression which is  the $\phi$ integral evaluated at its upper
limit $\phi_2$, where the region of $(x,y)$ integration is the one
given in Eq.~(14) {\em plus} its reflection about the line $x=a/2$,
(see Fig.~5).  This leads to the following expression for the
top/bottom terms:

\begin{equation}
 \int_{0}^{a}dx\int_{0}^{b}dy  \left[\int d\phi
\, \left(s(x,y,\phi) - \frac{s(x,y,\phi)^2}{2l}\right)
\right]_{\phi = \phi_2(x,y)},
\end{equation}
where $\phi_2(x,y) = \tan^{-1}(x/y)$ and $s(x,y,\phi)$ is given by
Eq.~(13).  One can go through the same logic for the scattering
contribution from the sides and it  can be seen, by making the
substitution $x \rightarrow y$ and $y \rightarrow x$, that the
expression obtained is identical to the one above with the exception
that the upper limits of the $x$ and $y$ integrals are reversed.
Summing these two contributions and performing the integrations we
arrive at the total expression for the case in which $l >\sqrt{a^2
+b^2}$,
\begin{eqnarray}
\tilde{l}_{ex} & = & \frac{a}{\pi}\ln\left(\frac{b +
\sqrt{a^2 +b^2}}{a}\right)
+ \frac{b}{\pi}\ln\left(\frac{a +
\sqrt{a^2 +b^2}}{b}\right) \nonumber\\
  &  & + \frac{1}{3\pi ab}\left[a^3 +b^3 -
(a^2 +b^2)^{3/2} \right] -\frac{ab}{2\pi l} \;\; \equiv \;\; H.
\end{eqnarray}

The symmetry considerations illustrated here are not essential for
performing the calculation above, but they become increasingly useful
as one examines more complicated cases.  Specifically, let us consider
the case when $l< a,b$.  As in the circular calculation, we will derive
the results for a limited portion of this range and then demonstrate
that they are applicable for the entire range.  We will initially
require $l$ to be less than $a/2$ and $b/2$.  Taking, once again, the
region of $(x,y)$ integration to be the lower-left quarter of the cross
section, one can  see that {\boldmath $l$} can only hit the bottom and
left sides.
 The task is, as before, to discern which regions of $(x,y,\phi)$ space
place us in the SH regime and which place us in the NH regime. The
simplest way of visualizing this is to consider a circle of radius $l$,
with its center located at various points $(x,y)$ within the lower-left
quarter, and to examine the manner in which it intersects the
boundaries.  We shall be particularly interested in the bottom wall, as
the sides can be related to it by symmetry arguments.  The relevant
regions are shown in Fig.~6.  In Region I, within a radius $l$ from the
bottom left corner, our circle  intersects the bottom wall exactly
once.  For points within this region, when one considers the
integration of $\phi$ over the bottom wall ({\em i.e.} from $\phi_1$ to
$\phi_2$), it can be seen that we are in the NH regime from $\phi_1$ to
some angle $\phi_1^*$ and in the SH regime from $\phi_1^*$ to
$\phi_2$.  $\phi_1^*$ is determined by the condition that
$s(x,y,\phi_1^*) = l$ and is given by
\begin{equation}
\phi_1^* = -\tan^{-1}\left(\frac{\sqrt{l^2-y^2}}{y} \right) .
\end{equation}
For points in region II, which have $y <l$ but are farther than $l$
from the corner, our circle  intersects the bottom wall in two places,
$\phi_1^*$ and $\phi_2^*$. $\phi_1^*$ is as defined above and
$-\phi_2^* = \phi_1^*$. In this region, we find as we integrate $\phi$
across the bottom wall, that we are in the SH regime for $\phi_1^* <
\phi < \phi_2^*$ and in the NH regime elsewhere.  Finally, in Region
III, where $y>l$, our circle does not intersect the bottom wall at all
and the NH expression applies to the entire $\phi$ integration.

Now let us consider the top wall.  It is clear that  the NH expression
applies as we integrate from $\phi_5$ to $\phi_6$ for origination
points anywhere in the lower left quarter of the cross section. We can
now apply the symmetry manipulations described for the $l > \sqrt{a^2
+b^2}$ case to the present scenario.  First, we rotate the ``top"
integration ($\phi_5$ to $\phi_6$)  by 180 degrees to convert it to an
integration over the upper-right quarter from $\phi_1$ to $\phi_2$.  We
can now combine this contribution with the integral obtained for region
III, as they both represent integrations from $\phi_1$ to $\phi_2$
where the NH expression holds for the entire range.  The various $\phi$
integrals we need to evaluate, then, are shown in Fig.~7a, with their
respective regions of $(x,y)$
 integration represented pictorially. These integrals represent the
 total contribution from integrations over the top and bottom walls to
 our expression for  $\tilde{l}_{ex}$, {\em i.e.,} the top/bottom
 terms.  At this stage we can divide up the $\phi$ integrals into their
 constituent upper and lower limit parts, and by performing the type of
 variable substitutions and accompanying $(x,y)$ integration region
 reflections discussed for the previous case
(see Appendix B),
 reduce the problem to that shown schematically in Fig.~7b.  The top/bottom
 contributions derived from Fig.7a are indicated by the diagrams on the
 left.  Those on the right represent the contributions from the sides.
 To illustrate how these schematics are converted to mathematical
 expressions, we will write out the top/bottom terms (those indicated
 by the  diagrams on the left in Fig.~7b) explicitly:
\begin{eqnarray}
\mbox{top/bottom}& = &\int_{0}^{l} dx
\int_{0}^{\sqrt{l^2-x^2}}dy \left[ \int d\phi \left(s(x,y,\phi)
 -\frac{s(x,y,\phi)^2}{2l}\right)
\right]_{\phi =\tan^{-1}(x/y)} \nonumber\\
 &  & +  \int_{0}^{l}dy \int_{\sqrt{l^2-y^2}}^{a}
 dx\left[\int d\phi \left(s(x,y,\phi)
-\frac{s(x,y,\phi)^2}{2l} - \frac{l}{2}\right)\right]_{\phi =
 \tan^{-1}(\sqrt{l^2 - y^2}/y)}  \nonumber\\
   &   & + \left[\int_{0}^{a}dx\int_{0}^{b}dy
 -\int_{0}^{l}dx\int_{0}^{\sqrt{l^2-x^2}}dy\right]\left[\int d\phi
 \frac{l}{2}\right]_{\phi =\tan^{-1}(x/y)},
\end{eqnarray}
where, again, $s(x,y,\phi)$ is given by Eq.~(13). An analogous
expression describes the side wall contributions.  The only changes are
in the limits of the $(x,y)$ integrations, and these are indicated by
the diagrams on the right in Fig.~7b.  Combining all contributions and
performing the integrations it can be shown that the total expression
for the case where $l <a,b$ is
\begin{equation}
\tilde{l}_{ex}  = \frac{l}{2} - \frac{(a +b)l^2}{3 \pi ab} +
\frac{l^3}{12\pi ab}
\equiv F.
\end{equation}

Recall that the results above were derived for the case in which $l <
a/2,b/2$.  The case where $b/2 < l <b$  appears, on the surface, to be
quite different, since, in an analysis of the type described in Fig.~6,
one would need to consider the possibility of hitting the top wall.
However, an examination of the forms of the integration regions
depicted in Fig.~7b reveals that they undergo no meaningful change
until $l$ becomes greater than $b$.  This is one of the benefits of
the  manipulations which result in Fig.~7b; they reveal the fundamental
equivalence of apparently disparate cases.  The other principle benefit
of the technique is that it obviates the need to extensively evaluate
the various SH and NH regions for other, {\em non-identical} cases
since the diagrams shown in Fig.~7b can be logically extended to other
regimes.   The case where $b < l < a$ for instance, is shown in Fig.~8a. We
note that the difference between the
 top/bottom diagrams and the side diagrams for this case results in an
 assymetrical dependence of $\tilde{l}_{ex}$ on $a$ and $b$. Performing
 the integrations indicated in the figure, we find that
 \begin{equation}
\tilde{l}_{ex}  = F + G(b),
\end{equation}
where,
\begin{eqnarray}
 G(q) &\equiv & -\frac{(q-l)^4}{12 \pi abl} + \frac{q}{\pi}\ln
 \left(\frac{l}{q} + \sqrt{(l/q)^2 - 1)} \right) \nonumber\\
&  & -\frac{l}{\pi}\cos^{-1}\left(\frac{q}{l}\right) + \frac{(l^2 -
 q^2)^{3/2}}{3 \pi ql}
\end{eqnarray}
and $F$ is as defined in Eq.~(19).

The final case to consider is that for which $a,b < l < \sqrt{a^2 +
b^2}$.  The appropriate diagrams are shown in Fig.~8b and they lead to
the result,
\begin{equation}
\tilde{l}_{ex}  = F + G(a) + G(b).
\end{equation}
One final remark: the case where $l > \sqrt{a^2 + b^2}$ which we
analyzed first can now be seen to result from the final possible
extension of the diagrams shown in Figs. 7 and 8.  The shaded regions
on the top line of Fig.~8b expand to encompass the entire rectangles
and the terms on the second line go to zero.

\medskip
\noindent
\begin{center}
{\bf III. BOUNDARY ORIGINATION}
\end{center}

All of the above expressions, for both the circular and rectangular
cases, were derived on the assumption of uniform origination, {\em
i.e.,} we assume that the excitation can originate anywhere within the
cross section with equal likelihood.  Recalling the physical
interpretation of the exchange length, however, we note that the
excitations we are considering are those that have just undergone a
scattering event within the cross section plane.  As pointed out in
Ref. \cite{flik}, when the mean-free path of the excitation becomes
much longer than the sample dimensions, the excitation becomes
increasingly likely to scatter on a boundary and our assumption of
uniform origination needs to be replaced with a boundary origination
description.   The expressions appropriate for boundary origination are
as follows:

\medskip
\noindent
Circular:
\begin{eqnarray}
\tilde{l}_{ex}  &= &\frac{2}{\pi} \int^{\pi/2}_0 d\phi \left(s(R,\phi) -
 \frac{s(R,\phi)^2}{2l}\right) \nonumber\\
  & = & \frac{4R}{\pi} - \frac{R^2}{l}
\end{eqnarray}

\medskip
\noindent
Rectangular:
\begin{eqnarray}
\tilde{l}_{ex} & = & \frac{4}{\pi(a+b)}\left\{\int^a_0 dx
\left[ 2\int^{\tan^{-1}(b/x)}_0 d\phi \, (s-s^2/2l) +
 \int^{\tan^{-1}(\frac{a-x}{b})}_{-\tan^{-1}(\frac{x}{b})} d\phi \,
 (s-s^2/2l)\right] \nonumber \right.\\
&  & \left. + \int^b_0 dy \left[ 2\int^{\tan^{-1}(a/y)}_0 d\phi \,
 (s-s^2/2l) + \int^{\tan^{-1}(\frac{b-y}{a})}_{-\tan^{-1}
(\frac{y}{a})} d\phi \, (s-s^2/2l)\right]\right\} \nonumber\\
& = & \frac{2}{\pi (a +b)} \left[\left( \frac{a^2}{2} + ab\right)
\ln\left(\frac{b + \sqrt{a^2 +b^2}}{a}\right)
+ \left( \frac{b^2}{2} + ab\right)\ln\left(\frac{a + \sqrt{a^2
+b^2}}{b}\right)\right] \nonumber\\
&  & + \frac{(a^2 +b^2)}{\pi(a +b)} -\frac{\sqrt{a^2 + b^2}}{\pi}
 -\frac{ab}{\pi l}
\;\; \equiv \;\; J
\end{eqnarray}

The appropriate procedure \cite{flik} is to match these solutions to
those for uniform origination at large $l$'s.   This can be
accomplished by a simple process which is as follows.  If we wish the
uniform solution, which we will call solution 1, to apply at some $l =
l^{\prime}$ and we wish the boundary solution, which we will call
solution 2, to apply for large $l$, the matched solution may be derived
by the expression,
\begin{equation}
\mbox{matched solution} = ({\rm sol}\; 1)
+ \exp(-ml^{\prime}/l)({\rm sol}\; 2 - {\rm sol}\; 1).
\end{equation}
where $m$ is a matching parameter.  We will take $l^{\prime}$ to be
$2R$ for the circular case and $\sqrt{a^2 + b^2}$ for the rectangular
case. The matched solutions then become

\bigskip
\noindent
Circular case:
\begin{equation}
 \tilde{l}_{ex}   =  \frac{8R}{3\pi} -\frac{R^2}{2l}   +
 \exp(-m2R/l)\left(\frac{4R}{3\pi} -\frac{R^2}{2l}\right).
\mbox{\hspace*{.5in}$l \geq 2R$}
\end{equation}

\bigskip
\noindent
Rectangular case:
\begin{equation}
\tilde{l}_{ex}  =  H +\exp(-md/l)[J-H]
\end{equation}
where $ d = \sqrt{a^2 + b^2}$, and $H$ and $J$ are defined by Eqs. (16)
and (24), respectively.  These forms result in small discontinuities at
$l= l^{\prime}$ when they are combined with the solutions for $ l <
l^{\prime}$ derived in the previous sections, Eqs. (11) and (22).  If
the matching parameter $m$ is chosen to be four, this discontinuity is
less than 1.5$\%$.  It should be noted that, for the rectangular case
when $a \gg b$,  we can expect to be largely in the boundary
origination regime before $l$ exceeds the length of the diagonal.  The
above equation does not allow for this possibility and is therefore
most applicable to cases in which $a$ is not too different from $b$.
For samples where one dimension is much larger than the other, we refer
the reader to the results of Ref. \cite{flik} where an expression for
thin films is derived.

\medskip
\noindent
\begin{center}
{\bf IV. DISCUSSION}
\end{center}

For the sake of clarity, the results for all the cases examined are
summarized in Table I.  Also included is the square subcase of the
rectangular expression.

\medskip
\noindent
I. Circular case:
\begin{eqnarray}
\mbox{$i.$ $l< 2R$} &  & \nonumber \\
\tilde{l}_{ex}  & = & \frac{8R}{3\pi} -\frac{R^2}{2l} +
\frac{l}{\pi}\cos^{-1}\left(\frac{l}{2R}\right) + \frac{R^2}{2\pi
l}\cos^{-1}\left(\frac{l^2}{2R^2} - 1 \right)   \nonumber \\
   &  &  - \frac{(4R^2-l^2)^{1/2}}{\pi}\left(\frac{13}{12} + \frac{l^2}{24R^2}
\right).\\
\mbox{$ii.$ $l \geq 2R$} &  & \nonumber \\
 \tilde{l}_{ex}  & = & \frac{8R}{3\pi} -\frac{R^2}{2l}   +
\exp(-8R/l)\left(\frac{4R}{3\pi} -\frac{R^2}{2l}\right).
\end{eqnarray}

\medskip
\noindent
I. Rectangular case:
\begin{eqnarray}
\mbox{$i.$ $l< a,b$} &  & \nonumber \\
\tilde{l}_{ex} & = &  F \\
\mbox{$ii.$ $b < l < a$} &  & \nonumber \\
\tilde{l}_{ex} & = & F + G(b)\\
\mbox{$iii.$ $a,b < l <d $ } &  & \nonumber \\
\tilde{l}_{ex} & = & F + G(a) + G(b)\\
\mbox{$iv.$ $l \geq d $ } &  & \nonumber \\
\tilde{l}_{ex} & = &  H +\exp(-md/l)[I-H]
\end{eqnarray}
where  $F$, $G$, $H$ and $I$ are defined in Eqs. (18), (20), (15), and (23),
respectively.

These expressions can now be used to aid in the calculation of
transport quantities in samples where boundary scattering is expected
to play a role.  We will examine the thermal transport in particular.
In the kinetic theory approximation, the thermal conductivity is given
by,
\begin{equation}
\kappa = \frac{1}{3} C v l,
\end{equation}
where $C$ is the contribution to the specific heat from the carrier in
question and $v$ is the carrier velocity.  Now, in Ref. \cite{flik}, it
is asserted that, within this approximation, the transport $\kappa_z$
along the $z$ axis in a sample of confined geometry can be obtained
from the bulk thermal conductivity $\kappa_{\infty}$ and the exchange
length along that axis by the expression
\begin{equation}
\kappa_z = \kappa_{\infty} \frac{\tilde{l}_{ex}}{l/2}
\end{equation}
It can easily be seen that this is equivalent to writing $\kappa_z$ as a
function of $l$ as
\begin{equation}
\kappa_z(l) = \kappa_{\infty}(l_{\em eff})
\end{equation}
with
\begin{equation}
l_{\em eff} = 2\tilde{l}_{ex}.
\end{equation}
In other words, the transport in the small sample can be obtained from
the expression for the bulk transport by replacing $l$ in the latter
expression with $l_{\em eff}$ as defined above.

Because of the simple geometric nature of this argument, it is
plausible that this sort of analysis can be applied to more
sophisticated treatments of the thermal conductivity as well.  For
instance, in the case of phonon transport,  the thermal conductivity is
often written in the Debye approximation as an integral over phonon
frequencies \cite{berman};
\begin{equation}
\kappa_p(T)={k_B\over 2\pi^2v}({k_B\over \hbar})^3 T^3
\int_0^{\theta_D/T} dx\,{{x^4\,e^x}\over (e^x-1)^2}\, \tau(T,x),
\end{equation}
where $x$ is the reduced phonon frequency $\hbar \omega /k_B T$, $k_B$
is the Boltzmann constant, $\theta_D$ is the Debye temperature and
$\tau(T,x)$ is the frequency dependent scattering time.  The total
inverse scattering time,  $\tau(T,x)^{-1}$, is usually expressed as a
summation of the inverse scattering times from scatterers of  various
types, {\em e.g.}, point defects, phonon-phonon umklapp processes,
etc.  Within this context, the effect of boundaries is typically
incorporated by adding a frequency independent term to this total of
the form \cite{casimir}
\begin{equation}
\tau_{b}^{-1} = v/\alpha d,
\end{equation}
where $d$ represents the sample dimension and $\alpha$
 is a geometrical factor.

We propose that greater accuracy may be achieved from calculations
involving Eq.~(32) by omitting the boundary scattering term in the
total inverse scattering time and utilizing the above equations for the
exchange length to modify the mean-free path instead.  The proposed
procedure is as follows:  the factor $\tau$ in Eq.~(32) can readily be
replaced by $l(x,T)/v$, at which point the integration can be seen to
be over the frequency dependent mean-free path multiplied by another
$x$ dependent factor.  For each such mean-free path, $l(x,T)$, a value
of $\tilde{l}_{ex}$ can be derived by using the equation appropriate
for the geometry of the particular sample under investigation.  Each
$l(x,T)$ in the integral can then be replaced by an $l_{\em eff}(x,T) =
2\tilde{l}_{ex}(x,T)$ as indicated in Eq.~(31).  The proper, boundary
limited value of the thermal conductivity is then obtained by
integrating over the $l_{\em eff}(x,T)$, with the other factors in
Eq.~(32) left unaltered.

The dependence of $l_{\em eff}$ on $l$ for various  geometries is shown
in Fig.~9.  Also shown in Fig.~9 is the result when $l_{\em eff}$ is
determined by the expression
\begin{equation}
l_{\em eff} = \left(\frac{1}{\alpha d} + \frac{1}{l}\right)^{-1},
\end{equation}
where $\alpha$ is taken to be 1 for the circular case and 1.12 for the
square case \cite{casimir}.  This expression results from making the
assumptions summarized in Eq.~(33), {\em i.e.,} the effect of boundary
scattering is addressed by adding a constant term to the total inverse
scattering time.  One can see that the discrepancy is approximately
20$\%$ in the region where $l$ is close to the sample dimensions.
While this error is not terribly large for the square and circular
cases, Eq.~(34) can not be used at all for the rectangular geometry,
where the sample can not be characterized by a single dimension.

\medskip
\noindent
\begin{center}
{\bf V. CONCLUSIONS}
\end{center}

In summary, we have presented analytical expressions for the effects of
boundary scattering in samples where the bulk carrier mean-free path is
determined by other scatterers present.  Results are derived for axial
transport in long, narrow samples of circular and rectangular cross
section, where scattering at the boundaries is diffuse.  The results
are incorporated into a definition of an effective mean-free path for
axial transport which can be used to calculate coefficients such as the
thermal conductivity.  Though we have focused on thermal transport in
the present work, the expressions derived here could be of use in the
examination of a variety of transport phenomena in confined
geometries.

\medskip
\noindent
\begin{center}
{\bf ACKNOWLEDGMENTS}
\end{center}

FN acknowledges partial support from a GE fellowship,
 a Rackham grant, the NSF through grant DMR-90-01502, and SUN Microsystems.

\medskip
\noindent
\begin{center}
{\bf APPENDIX A: Evaluation of Eq.~(10)}
\end{center}

The evaluation of the integral expressed in Eq.~(10) is not entirely
straightforward and, for that reason,  it will be discussed in greater
detail here.
  The $\rho$ integration is easily performed, resulting in the
  expression
\begin{equation}
\mbox{``s term"} = \frac{2}{\pi R^2}  \left. \int_{0}^{\phi_2^*}d\phi \, \,
A(\rho,\phi)\right|^R_{\sqrt{R^2 - l^2\sin^2\phi} - l\cos\phi} ,
\end{equation}
where
\begin{equation}
A(\rho,\phi) =\int d\rho \rho \left( s- \frac{s^2}{2l}\right)
 = -\frac{(R^2 - \rho^2\sin^2\phi)^{3/2}}{3\sin^2\phi} -
 \frac{\rho^3}{3}\cos\phi .
\end{equation}
We have used Eq.~(6) for $s(\rho,\phi)$ and written out the lower limit
$\rho^*$ explicitly.  The difficulty lies in the complicated nature of
this lower limit.  Simply substituting it into $A(\rho,\phi)$ leads to
a result which is not easily dealt with.  However, we can make use of
other information to work this expression into a more tractable form.
Specifically, we know that, from the definition of $\rho^*$, when $\rho
= \rho^*$, $s = l$.  Therefore, if  $A(\rho,\phi)$ can be expressed in
terms of $s$,  then, when the lower limit $\rho^*$  is substituted in
for $\rho$, we can also substitute $l$ for $s$.  After some algebraic
manipulations, $A(\rho,\phi)$ can be expressed as
\begin{equation}
A(\rho,\phi) = \frac{1}{\sin^2\phi}\left(\frac{2\rho^3
\cos \phi}{3} + \rho^2 s \cos^2 \phi - \rho R^2 \cos \phi - s^3/3\right).
\end{equation}
which, upon substitution of $\rho^*$ for $\rho$ and $l$ for $s$, becomes
\begin{equation}
A(\rho^*,\phi) = \frac{2l^3\cos^2\phi}{3} - \frac{l^3}{3} -
\frac{\cos \phi}{3} \sqrt{R^2 - l^2\sin^2\phi}\left(2l^2
+ \frac{R^2}{\sin^2\phi}\right).
\end{equation}
This can now be integrated with respect to $\phi$, and, when combined
with the upper limit term (which presents no special problems) gives
the result
\begin{equation}
\mbox{``s term"} = \frac{8R}{3 \pi} - \frac{(4R^2 -l^2)^{3/2}}{3 \pi R^2}.
\end{equation}

\medskip
\noindent
\begin{center}
{\bf  APPENDIX B: Derivation of Fig.~7b}
\end{center}

In this appendix, we will elucidate the details of the process by which
one goes from the form expressed in Fig.~7a to that shown in Fig.~7b.
To aid in this discussion, a key of relevant diagrams, each
representing a region of integration in the $(x,y)$ plane, is provided
in Fig.~10. We will also abbreviate the expression $s - s^2/2l$ as
$W(s)$.

We start by noting that the first term in square brackets on the first
and second lines of Fig.~7a is the same. Therefore, in evaluating these
particular two terms, their regions of $(x,y)$ integration, $I_1$ and
$I_2$, can be combined to give region $I_7$, (see Fig.~10).  The next
step, as discussed in the main text, is to separate each integral into
its upper and lower limit parts and, by performing a series of variable
substitutions, reduce the problem to the smallest possible number of
distinct terms. As an example of this dividing and substituting
process, let us examine the term on the third line of Fig.~7a. This
term can be written, with the help of Fig.~10, as $$I_3 \times
\left[\int_{\phi_1}^{\phi_2} d\phi \, \frac{l}{2}\right]$$ where the
multiplication notation indicates the double integration of the term in
brackets over the $(x,y)$ region defined as $I_3$.  Now consider the
lower limit, $\phi_1 = -\tan^{-1}\left(\frac{a-x}{y}\right)$.  If we
make the change of variables $x^{\prime} = a-x$, it can easily be seen
that
\begin{equation}
-I_3 \times \int d\phi \,\left.  \frac{l}{2}\right|_{\phi =
-\tan^{-1}\left(\frac{a-x}{y}\right)} = I_6 \times  \int d\phi \,\left.
  \frac{l}{2}\right|_{\phi = \tan^{-1}\left(\frac{x^{\prime}}{y}\right)}.
\end{equation}
This variable change results, then, in a term that looks very much like
the expression that one gets for the upper limit $\phi_2 =
\tan^{-1}\left(\frac{x}{y}\right)$, with the exception that the region
of $(x,y)$ integration, $I_6$, is the mirror reflection of the original
region, $I_3$, about the line $x=a/2$.  This procedure will now be
applied to the terms in Fig.~7a as follows:
\begin{eqnarray}
I_1 \times \int_{\phi_1^*}^{\phi_2} d\phi \, W(s) & \Longrightarrow &
 I_1 \times  \left. \int d\phi \,  W(s)\right|_{\phi = \phi_2} +
 I_4 \times \left. \int d\phi \,  W(s)\right|_{\phi = \phi_2^*} \\
I_2 \times \int_{\phi_1^*}^{\phi_2^*} d\phi \, W(s) & \Longrightarrow &
 I_2 \times \left. \int d\phi \,  W(s)\right|_{\phi = \phi_2^*} +
 I_5 \times \left. \int d\phi \,  W(s)\right|_{\phi = \phi_2^*} \\
I_2 \times \int_{\phi_2^*}^{\phi_2} d\phi \,  \frac{l}{2} & \Longrightarrow &
 I_2 \times  \int d\phi \,\left.   \frac{l}{2}\right|_{\phi = \phi_2} -
 I_2 \times  \int d\phi \,\left.   \frac{l}{2}\right|_{\phi = \phi_2^*} \\
I_3 \times \int_{\phi_1}^{\phi_2} d\phi \,  \frac{l}{2} & \Longrightarrow &
 I_3 \times  \int d\phi \,\left.   \frac{l}{2}\right|_{\phi = \phi_2} +
 I_6 \times  \int d\phi \,\left.   \frac{l}{2}\right|_{\phi = \phi_2} \\
I_7 \times \int_{\phi_1}^{\phi_1^*} d\phi \,  \frac{l}{2} & \Longrightarrow &
 I_8 \times  \int d\phi \,\left.   \frac{l}{2}\right|_{\phi = \phi_2} -
 I_8 \times  \int d\phi \,\left.   \frac{l}{2}\right|_{\phi = \phi_2^*}.
\end{eqnarray}
In the above we have made use of the fact that $\phi_2^* = -\phi_1^*$,
with $\phi_1^*$ given by Eq.~(17).  We note that one of the integrals
had the change of variables applied to both limits (Eq. 45) and one did
not have it applied at all (Eq. 43) so as to generate the fewest number
of similar terms.  This number can be now be seen to be four: those
involving $W(s)$, evaluated at $\phi_2$ and $\phi_2^*$, and those
involving $l/2$, evaluated at these same two limits.  The final step in
the derivation of Fig.~7b is the combining of terms, wherein the $I$'s
of like expressions are patched together like pieces of a puzzle.  This
process is illustrated in Fig.~11 for the terms involving $l/2$
evaluated at $\phi_2$.  It can be seen that four contributing diagrams,
$I_2$, $I_3$, $I_6$, and $I_8$, combine to yield $I_9$, which is the
form that appears in Fig.~7b.  The other top/bottom terms in Fig.~7b
are derived in a similar manner.

\newpage
\noindent
\begin{large}
{\bf Figure Captions}
\end{large}

\noindent
Figure 1:  Schematic showing the geometry relevant for the calculation
of the exchange length for a wire of circular cross section.  The case
pictured is that for which {\boldmath $l$} hits the boundary.  Note
however that the quantity $s$ is always defined as the segment within
the cross section plane that stretches from the origination point to
the boundary, regardless of whether or not {\boldmath $l$} actually
hits the boundary.

\bigskip
\noindent
Figure 2:  Schematic of the various regimes when $l < R$  and $R-l <
\rho < R$.  The angle $\phi^*$ is defined by the condition that
$s(\rho, \phi^*) = l$.  For angles greater than  $\phi^*$,  {\boldmath
$l$} never hits the boundary.  Arrows indicate possible orientations of
the vector {\boldmath $l$}  when $\theta = \pi/2$.

\bigskip
\noindent
Figure 3:  a) Schematic indicating  how the angle $\phi_2^*$, above
which  {\boldmath $l$} never hits the boundary for any $\rho$, is
determined.  When $\phi = \phi_2^*$,  $s(R, \phi_2^*) = l$. b)
Schematic indicating how
 $\rho^*$, above which  {\boldmath $l$} sometimes hits the boundary, is
 determined for a given $\phi$.   When $\rho = \rho^*$,  $s(\rho^*,
 \phi) = l$.   Arrows indicate possible orientations of the vector
 {\boldmath $l$} when $\theta = \pi/2$ for varying $\rho$ and fixed
 $\phi$.  The relevant region for the $\rho$ integration portion of the
 $s$ term is indicated by the bracket.

\bigskip
\noindent
Figure 4:  Diagram indicating the various values of $\phi$ which are
used in the determination of the exchange length in a sample of
rectangular cross section.

\bigskip
\noindent
Figure 5:  Schematic representation of the symmetry processes used to
simplify the calculation of the average exchange length.  The
rectangular diagrams represent double integrals in the $(x,y)$ plane
over the shaded regions.

\bigskip
\noindent
Figure 6:  Diagram indicating the various regions in the $(x,y)$ plane
which must be treated in  the calculation of the average exchange
length.  In region I, a circle of radius $l$ intersects the bottom wall
once.  In region II, a circle of radius $l$ intersects the bottom wall
twice and in region III there is no intersection.  The curve in the
figure has radius $l$.

\bigskip
\noindent
Figure 7: a) Schematic representation of the various terms involved in
the calculation of the contribution from the top and bottom walls to
the exchange length for the case in which $l < a/2, b/2$.  The
rectangular diagrams represent integrations in the $(x,y)$ plane over
the shaded regions.  b)  Terms used to calculate the average exchange
length $\tilde{l}_{ex}$ after symmetry manipulations have been
applied.  The leftmost diagram in each term represents the contribution
from the top and bottom walls and the rightmost diagram represents the
contribution from the side walls.  These terms must be multiplied by a
pre-factor to yield the final result for $\tilde{l}_{ex}$. The curves
in the diagrams have radius $l$ and are centered at the lower left
corner.

\bigskip
\noindent
Figure 8: a) Terms used in the calculation of the average exchange
length for the case in which $b < l < a$.  The diagrams representing
the $(x,y)$ integrations can be seen to be logical extensions of the
previous case.  Again, the curves have radius $l$. b) Terms used in the
calculation of the average exchange length for the case in which $b,a <
l < \sqrt{a^2 + b^2}$.

\bigskip
\noindent
Figure 9:  a)  Plot of the reduced effective mean-free path  versus the
reduced bulk mean-free path for the circular case.  The solid line
indicates  the results when the exchange length method is used while the
dashed line indicates Casimir's results using Eq.~(34) with $\alpha = 1$.
The inset shows a blow up of the low $l$ region and has the same axes
as the surrounding plot.  b)  Results for the rectangular case.  The
solid line shows the result for $a = b$ and the dot-dashed line
represents the case in which $a = 2b$.  As in a), the dashed line shows
the result from Eq.~(34) with $\alpha = 1.12$. Inset is as in a).

\bigskip
\noindent
Figure 10: Key of diagrams used in Appendix B for the derivation of
Fig.~7b.  Curves in the diagrams have radius $l$.

\bigskip
\noindent
Figure 11:  Illustration of the way in which the various integration diagrams
 for the terms involving $l/2$ evaluated at $\phi_2$ combine to a single
diagram.

\newpage
\noindent
\begin{large}
{\bf Table Caption}
\end{large}

Table I:  Summary of the equations for  $\tilde{l}_{ex}$ for the
various different geometries and the various ranges of the mean-free
path $l$. $R$ is the radius  for the circular case, and $a$ and $b$ are
the lengths of the sides for the rectangular case, with $b$ taken to be
the shorter of the two.  $d$ is equal to $\sqrt{a^2 + b^2}$.


\newpage

\renewcommand{\arraystretch}{2}
\oddsidemargin -0.2cm
\textwidth 15.9cm
\topmargin -0.3cm
\textheight 20.0cm

\begin{tabular}{||c|c|c||} \hline
Cross Section  & Range  &  $\tilde{l}_{ex}$ \\
\hline \hline
Circle   &  $l < 2R$ & $\begin{array}{l}
\frac{8R}{3\pi} -\frac{R^2}{2l} +
\frac{l}{\pi}\cos^{-1}\left(\frac{l}{2R}\right) + \frac{R^2}{2\pi
l}\cos^{-1}\left(\frac{l^2}{2R^2} - 1 \right) \\
 - \frac{(4R^2-l^2)^{1/2}}{\pi}\left(\frac{13}{12} + \frac{l^2}{24R^2} \right)
\end{array}$\\
\cline{2-3}
    &  $ l \geq 2R$  &  $\frac{8R}{3\pi} -\frac{R^2}{2l}   +
\exp(-8R/l)\left(\frac{4R}{3\pi} -\frac{R^2}{2l}\right)$\\
\hline
Rectangle  & $l < b$ & $F(a,b)$ \\ \cline{2-3}
           & $b \leq l < a$ & $F(a,b) + G(b,a)$\\ \cline{2-3}
           & $a \leq l < d$  &   $F(a,b) + G(a,b) + G(b,a)$\\ \cline{2-3}
           & $ l \geq d$ & $ H +\exp(-4d/l)(J-H)$ \\
\hline
Square     & $l < a$  & $F(a,a)$ \\  \cline{2-3}
           & $a \leq l < \sqrt{2}a$ &  $F(a,a) + 2G(a,a)$ \\  \cline{2-3}
           & $l \geq \sqrt{2}a$     &  $\begin{array}{l}
\frac{a}{\pi}\left[2 + \exp(-\frac{4\sqrt{2}a}{l})\right]
\left[\ln(1 + \sqrt{2}) + (1-\sqrt{2})/3\right] \\
-\frac{a^2}{2\pi l}\left[1 + \exp(-\frac{4\sqrt{2}a}{l})\right]
\end{array}$ \\
\hline
\hline
\multicolumn{3}{|c|}
{$F(p,q) \equiv \frac{l}{2} - \frac{(p +q)l^2}{3 \pi pq} + \frac{l^3}{12\pi
pq}$}\\
\multicolumn{3}{|c|}
{$ G(p,q) \equiv  -\frac{(p-l)^4}{12 \pi pql} + \frac{p}{\pi}\ln
\left(\frac{l}{p} + \sqrt{(l/p)^2 - 1)} \right)
-\frac{l}{\pi}\cos^{-1}\left(\frac{p}{l}\right) + \frac{(l^2 - p^2)^{3/2}}{3
\pi pl}$}\\
\multicolumn{3}{|c|}
{$H \equiv \frac{a}{\pi}\ln\left(\frac{b + d}{a}\right)
+ \frac{b}{\pi}\ln\left(\frac{a + d}{b}\right)
  + \frac{1}{3\pi ab}\left[a^3 +b^3 -d^3 \right] -\frac{ab}{2\pi l}$}\\
\multicolumn{3}{|c|}
{$J  \equiv  \frac{2}{\pi (a +b)} \left[\left( \frac{a^2}{2} + ab\right)
\ln\left(\frac{b + d}{a}\right)
+ \left( \frac{b^2}{2} + ab\right)\ln\left(\frac{a + d}{b}\right)\right]
 + \frac{d^2}{\pi(a +b)} -\frac{d}{\pi} -\frac{ab}{\pi l}$} \\ \hline
\end{tabular}

\end{document}